\documentclass[twoside,fleqn]{article}
\usepackage{amssymb}
\usepackage{npb}
\usepackage{graphics}
\def\beq{\begin{eqnarray}}    
\def\eeq{\end{eqnarray}}      

\title{Gravitational waves in an anomaly-induced inflation}

\author{J.C. Fabris 
\address{Universidade Federal do Esp\'{\i}rito Santo, 
Departamento de F\'{\i}sica, Esp\'{\i}rito Santo, Brazil}, 
\qquad
A.M. Pelinson ,
\quad 
I.L. Shapiro
\quad 
and 
\quad
F.I. Takakura
\address{Departamento de Fisica, ICE, Universidade Federal de Juiz de
Fora, MG, Brazil}}
       
\begin{document}

\begin{abstract}
The behaviour of gravitational waves in the anomaly-induced inflationary 
phase is studied. The metric perturbations exhibit a stable behaviour, 
with a very moderate growth in the amplitude of the waves. The spectral 
indice is computed, revealing an almost flat spectrum.
\vspace{1pc}
\end{abstract}

\maketitle

The goal of the present study is to analyze the fate of gravitational 
waves in a background defined by an inflationary scenario created by a 
trace anomaly induced
by quantum effects generated by matter fields in the primordial 
Universe \cite{ana}. This model
is a generalization of the Starobinsky model \cite{staro}, where a 
similar mechanism has been developed. The counterterms
necessary to avoid divergences in the quantization of these conformal 
quantum fields in
a Friedmann-Robertson-Walker background leads to a higher derivative 
action with a trace anomaly. Details of the model may be found in 
\cite{ana}. An important point concerns
the fact that an inflationary phase may be obtained with this effective 
gravitational
action. When all fields are massless, this inflationary phase is an 
eternal de Sitter phase,
where many of the traditional problems of the inflationary scenario 
appear. However,
if massive conformal fields are included, the inflationary phase 
has more complicated
form. A detailed analysis of the massive case reveals that a 
huge amplification
of the scale factor may be obtained, of the order of ${10}^{10}$ 
e-folds. However,
from the observational point of view, only the last $65-70$ e-folds 
are relevant.
During this final period, the behaviour of the scale factor may be 
approximated by an
exponential function, the Hubble parameter being constant. It is 
possible that a transition
to the FRW standard scenario may be achieved solving the graceful 
exit problem \cite{shapiro}.
\par
The equation governing the behaviour of gravitational waves in the 
anomaly-induced model, in terms of the cosmic time $t$, is \cite{flavio}
$$
b_0\stackrel{....}{h} + b_1\stackrel{...}{h} 
+ b_2\ddot h + b_3\dot h + b_4h 
$$
\begin{equation}
+ n_1\frac{\nabla^2\dot h}{a^2} 
+ n_2\frac{\nabla^2\ddot h}{a^2} 
+ n_3\frac{\nabla^4h}{a^4}\,=\,0\,,
\label{1}
\end{equation}
with
$$ 
b_0 = \tilde b_0 \,, \quad 
b_1 = H\tilde b_1\,, \quad 
b_2 = H^2\tilde b_2\,, 
$$$$
b_3 = H^3\tilde b_3 \,, \quad 
b_4 = H^4\tilde b_4 \,, 
$$
\begin{eqnarray}
n_1 = H\tilde n_1\,, \quad n_2 = \tilde n_2 \,,
\quad n_3 = \tilde n_3 \,.
\end{eqnarray}
Here the quantities with tildes are pure numbers:
\begin{eqnarray}
\tilde b_0 = 1  \,, \quad 
\tilde b_1 = 6  \,, \quad 
\tilde b_2 = 11 \,, \quad 
\tilde b_3 = 6  \,, 
\nonumber 
\\ 
\tilde n_1 = - 2 \,,\quad 
\tilde n_2 = -2  \,,\quad 
\tilde n_3 = 1 \,,
\end{eqnarray}
while $\tilde b_4$ depends on the multiplet content 
of the matter fields:
\begin{equation}
b_4 = \frac{12}{(4\pi)^2}\biggr(\frac{N_0}{360} + 
\frac{11N_{1/2}}{360} + \frac{31N_1}{180}\biggl) \,,
\end{equation}
where $N_0$, $N_{1/2}$ and $N_1$ are the scalar, 
fermionic and vectorial numbers of
matter fields. For example, in the minimal standard model, 
$N_0 = 4$, $N_{1/2} = 24$ 
and $N_1 = 12$, leading to $b_4 \sim 0.2$.

As usual, $H = {\dot a}/{a}$ is
the Hubble parameter. The parameters described above take 
those value in the
last $65$ e-fold. In this case, the background model enters 
in a quasi-de Sitter and
the scale factor behaves essentially as
\begin{equation}
a \sim e^{Ht} 
\end{equation}
where, in an appropriate GUT model, 
$$
H \approx 10^{-5}M_{Pl}\,.
$$

Changing to the conformal time $\eta$ 
such that $dt = ad\eta$, the scale factor takes the form
\begin{equation}
a \sim - \frac{1}{H_0\eta}
\end{equation}
with $\eta < 0$. The Universe expands as $\eta \rightarrow 0_-$. 
In terms of this new time parameter, and performing a Fourier decomposition
 of the function $h = h(x,\eta) = h(\eta)e^{i\vec k.\vec x}$,
$\,k$ being
the wavenumber of the perturbation, we obtain the following fourth order 
differential equation describing the evolution of gravitational waves.
\begin{equation}
h^{iv} + 2k^2h'' 
+ \biggr\{\frac{\tilde b_4}{\eta^4} + k^4\biggl\}h = 0 \,.
\label{11}
\end{equation}
The units are such that $k = 1$ implies a comoving wavelength of the 
perturbation of the order of the Planck's length. Let us notice the
remarkably simple form of the equation (\ref{11}), which is essentially 
simpler that the equivalent equation in terms of the physical time 
(\ref{1}). 

We first investigate the behaviour of $h$ in the 
two asymptotic regimes: small and large wavenumbers.
For large wavenumbers $k >> 1$ and we may approximate 
the equation to
\begin{equation}
h^{iv} + 2k^2h'' + k^4h = 0 \,,
\end{equation}
whose solution is
\begin{equation}
h = c_\pm e^{\pm ik\eta} + c'_\pm\eta e^{\pm ik\eta} \,.
\end{equation}
Since $\eta$ approaches zero as the universe expands, 
the solution above is a combination 
of oscillatory and decreasing oscillatory modes. 
A quantum state can be implemented in the initial state 
(when $\eta \rightarrow - \infty$),
if, for example, $c_- = c_0/\sqrt{2k}$, $c_0 =$ constant, 
$c_+ = c'_\pm = 0$.
For small wavenumbers, on the other hand, the equation 
simplifies to
\begin{equation}
h^{iv} + \frac{\tilde b_4}{\eta^4}h = 0 \,.
\end{equation}
We can look for solutions under the form
$h \propto \eta^p$, where $p$ is a number. In this case,
$p$ obeys the algebraic relation
\begin{equation}
p(p - 1)(p - 2)(p - 3) + \tilde b_4 = 0 \,.
\end{equation}
There is no root for $p$ with $Re\,p < 0$. 
Hence, there is only decreasing modes in the
long wavelength approximation. 
\par
All the behaviours sketched above were confirmed through 
numerical integration. 
The results indicate that the inflationary phase in this model 
is quite stable with
respect to tensorial perturbations. There is a small production 
of gravitons
during the inflationary phase. This seems to be a positive 
result since a very large
production of graviton during the inflationary phase would lead 
to undesirable consequences
since a large amplification
of gravitational waves brings the problem of back reaction 
\cite{bran} and, at same time, renders the
linear approximation doubtful. No such problems exist in the 
behaviour of gravitational
waves in the context of the anomaly-induced inflationary scenario.

It is instructive to compare the model described above with 
the results obtained with the case of the standard inflation
based on a cosmological constant term (or an evolving scalar 
field) \cite{kolb}. Since a de Sitter
phase is a good approximation to the standard case, the 
scale factor behaves as before.
In this case the equation governing the behaviour of gravitational 
waves takes
the form
\begin{equation}
\label{eogw}
h'' - 2\frac{a'}{a}h' + \biggr\{k^2 - 2\biggr[\frac{a''}{a} 
- \biggr(\frac{a'}{a}\biggl)^2\biggl]\biggl\}h = 0 \,.
\end{equation}
The solution for this equation is
\begin{equation}
\label{in-sol}
h = \frac{1}{\sqrt{\eta}}\biggr\{c_1H^{(1)}_{3/2}(k\eta) 
+ c_2H^{(2)}_{3/2}(k\eta)\biggl\}\,,
\end{equation}
where the $H_\nu^{(1,2)}$ are Hankel functions of first and 
second kind, and the $c$'s are constants.

From the point of view of gravity production, directly 
connected with the amplification of
the gravitational waves, the anomaly-induced inflationary model 
is very different from
the standard one. While the solutions (\ref{in-sol}) exhibit 
an amplification
of the order $10^{65}$ in the last $65$ e-folds, in the 
anomaly-induced inflation
the amplitude of the gravitational waves remains essentially stable.
The huge amplification of gravitational waves in the 
standard inflationary
scenario brings essentially two problems: first, the non-linear 
regime is reached very quickly
unless the initial perturbations are excessively small; moreover, 
it is difficult to ignore
in this situation the back-reaction. The firsts problem
may be coped with through the quantum normalization: the initial 
amplitude must be
of the order of the inverse Planck's mass. The second one is object of
many investigations today \cite{bran}. These problems are absent
in the anomaly-induced inflationary model: since a perturbation 
originated in the beginning
of the de Sitter phase would have been amplified at the end
 of this phase by a factor of
order of unity, the perturbation enters in the radiative 
phase very small.
During the radiative phase,
the equation governing the evolution of gravitational waves 
is given by (\ref{eogw}),
with $a \propto \eta$ and with $a \propto \eta^2$ during the
 matter dominated phase.
The gravitational waves will be considerably amplified only 
during these phases. The non
linear regime may be attained only quite recently.
\par
We turn now to the problem of evaluation of the spectral 
indice of the power spectrum in
the anomaly-induced inflationary scenario. The power 
spectrum is defined as
\begin{equation}
\label{si}
P^2(k) = k^3\delta_k^2 \,.
\end{equation}
The spectral indice is obtained through the relation
\begin{equation}
n_T = \frac{d\ln P_k^2}{d\ln k} \,,
\end{equation}
where the subscript $T$ indicates that we are evaluating the 
power spectrum for the tensorial
modes. A scale invariant spectrum is characterized by the fact 
that the amplitude of the perturbation are the same for any 
value of $k$ when the perturbation cross the horizon.
This implies $n_T = 0$, when the expression (\ref{si}) is 
evaluated at the horizon crossing.

A fundamental point in the analysis to be performed is the 
question of initial conditions.
In the anomaly-induced inflation, as in the standard 
inflaton-based inflation, the seeds
of the initial perturbations are quantum mechanical. 
The normalized initial quantum spectrum takes the form
\begin{equation}
\label{ic}
h(\eta)_k \sim \frac{e^{-ik\eta}}{\sqrt{2k}} \quad .
\end{equation}
\par
Let us
consider the initial time $\eta_i \sim 1$ 
in Planck's unities. The final time is $\eta_f \sim
10^{-30}$. Suppose also that the inflationary 
phase ends at about $t \sim 10^{-38}\,s$,
consistent with the fact that $H_0 \sim 10^{-5}M_{Pl}$.
Since in our unities $k \sim 1$ means a initial 
perturbation of the order of the Planck's scale,
perturbations today in range from the Hubble horizon and 
some hundreds of megaparsecs, evaluated
today, implies initial perturbations with 
$2\pi10^{-3} < k < 2\pi10^{-1}$.
We now compute numerically the perturbations using the initial 
conditions (\ref{ic}) for the traditional inflationary scenario 
and for the anomaly-induced inflation. In both cases,
the spectrum is essentially flat: $n_T \sim 0$ at the end of the 
inflationary phase. 
Considering, for example, $\pi/100 < k < \pi/10$,
 the spectral indice is $n_T \sim -0.027$.
This may be compared with the spectral indice for the standard 
inflationary scenario which,
for a de Sitter phase, is strictly zero in the large wavenumber 
approximation.

\end{document}